\documentclass{article}
\usepackage{spconf,amsmath,graphicx}
\usepackage{booktabs}
\usepackage{color}
\usepackage{pifont}
\usepackage{wasysym}
\usepackage[justification=centering]{caption}

\usepackage{enumitem}
\setlist{nosep, leftmargin=14pt}

\usepackage{mwe} 

\title{DIGEST: Deeply supervIsed knowledGE tranSfer neTwork learning for brain tumor segmentation with incomplete multi-modal MRI scans
}
\name{Haoran Li$^{1, 2}$, Cheng Li$^{1}$, Weijian Huang$^{1,2,3}$,Xiawu Zheng$^{3}$,Yan Xi$^{1}$, Shanshan Wang$^{1, 3, 4}$$^{*}$}

\address{
$^{1}$Paul C. Lauterbur Research Center for Biomedical Imaging, Shenzhen Institutes of \\
Advanced Technology, Chinese Academy of Sciences, Shenzhen, Guangdong, China\\
$^{2}$University of Chinese Academy of Sciences, Beijing, China\\
$^{3}$Peng Cheng Laboratory, Shenzhen, Guangdong, China\\
$^{4}$Guangdong Provincial Key Laboratory of Artificial Intelligence in Medical Image \\ 
Analysis and Application, Guangdong Provincial People’s Hospital, \\
Guangdong Academy of Medical Sciences, Guangzhou, Guangdong, China\\
}
\begin{document}
%
\maketitle 
\begin{abstract}
Brain tumor segmentation based on multi-modal magnetic resonance imaging (MRI) plays a pivotal role in assisting brain cancer diagnosis, treatment, and postoperative evaluations. Despite the achieved inspiring performance by existing automatic segmentation methods, multi-modal MRI data are still unavailable in real-world clinical applications due to quite a few uncontrollable factors(e.g. different imaging protocols, data corruption, and patient condition limitations), which lead to a large performance drop during practical applications. In this work, we propose a Deeply supervIsed knowledGE tranSfer neTwork (DIGEST), which achieves accurate brain tumor segmentation under different modality-missing scenarios. Specifically, a knowledge transfer learning frame is constructed, enabling a student model to learn modality-shared semantic information from a teacher model pretrained with the complete multi-modal MRI data. To simulate all the possible modality-missing conditions under the given multi-modal data, we generate incomplete multi-modal MRI samples based on Bernoulli sampling. Finally, a deeply supervised knowledge transfer loss is designed to ensure the consistency of the teacher-student structure at different decoding stages, which helps the extraction of inherent and effective modality representations.
Experiments on the BraTS 2020 dataset demonstrate that our method achieves promising results for the incomplete multi-modal MR image segmentation task.
\end{abstract}
\section{Introduction}
\label{sec:intro}
Brain tumors are one of the leading causes of death in the world. Segmenting the tumor regions in multi-modal magnetic resonance (MR) images, which can provide multi-contrast imaging information, is a critical step in the clinical workflow for treatment planning and postoperative monitoring of patients with brain tumors. However, owing to the structural complexities and  histologic differences of brain tumors\cite{bakas2017advancing}, manually delineating the tumor boundaries in multi-modal MR scans is time-consuming and error-prone. Consequently, automated brain tumor segmentation methods are highly desired. Inspired by the success of deep learning(DL) in various fields, many DL-based medical image analysis methods have been developed for different application scenarios\cite{zhou2019d}\cite{wang2021annotation}, among which those for brain tumor segmentation based on multiple MR imaging (MRI) modalities are widely investigated. Despite the achieved promising results, it is not always feasible to have the complete modalities in real-world clinical applications due to different reasons, such as different imaging protocols, data corruption, and patient condition limitations. Since the performance of these methods highly depends on the modality integrity, issues may occur when deploying them directly in applications.

\begin{figure*}[htb]
    \centering
    \includegraphics[width=0.9\textwidth]{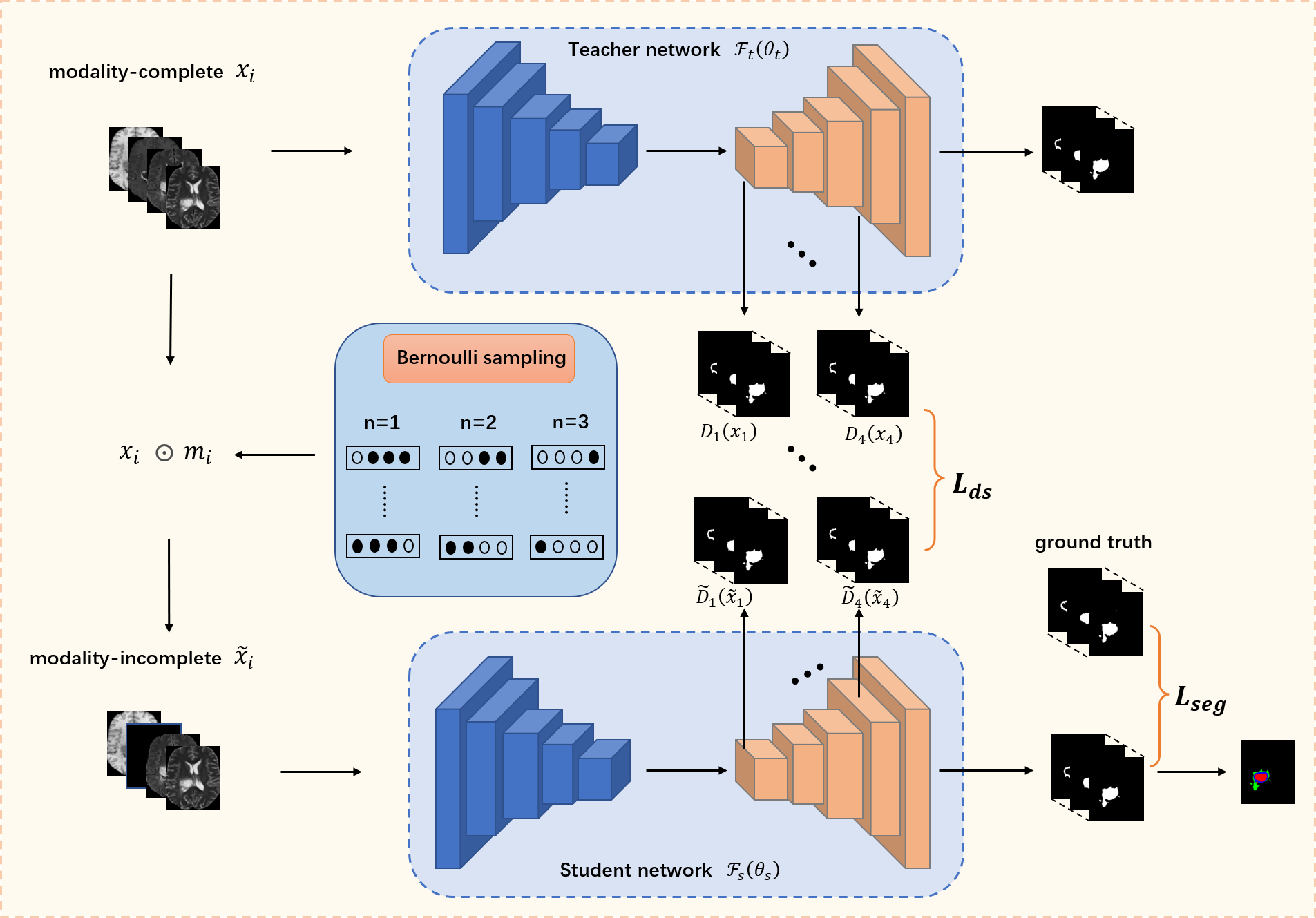}
    \caption{The proposed DIGEST. Deep supervision loss $L_{ds}$ constrain the generated auxiliary segmentation maps for each tumor sub-region at every decoding stage.}
    \label{fig:my_label}
\end{figure*}

To solve the missing modality problem, an intuitive solution is to train respective segmentation models for any subset of the complete MRI modalities. This approach has two limitations. On the one hand, models trained with limited modalities are less effective than those trained with the complete modalities. On the other hand, training multiple models is not elegant and flexible, and it confuses the end-users. To this end,  some methods are developed trying to synthesize the missing modalities by using support vector machines, random forests with regression strategy\cite{jog2017random}\cite{hofmann2008mri}. Unfortunately, it has been reported that the synthesized modality resulted in negligible performance improvement\cite{tulder2015does}. Recently, some works have built a shared latent space to search for the common representations over different MRI modalities and maintained the brain tumor segmentation performance of incomplete modality inputs by exploiting these representations. Havaei M et al.\cite{havaei2016hemis} has proposed a model named HeMIS which learns to embed the input data of each modality into the same latent vector space. The mean and variance of the input feature are computed to characterize the multi-modal information.
Followed by this work, Dorent R et al.\cite{dorent2019hetero} introduces a method named U-HVED, which utilizes the 3D-Unet with a Variational Auto Encoder (VAE) for multi-modal data to complete the missing modalities.

Existing methods have provided important insights for utilizing incomplete multi-modal MRI data. Further attempts are needed to find a more effective way of common feature extraction across multiple missing hetero-modal samples. Additionally, building explicit relationships between the common features and the modality-complete features is also important for the performance enhancement of brain tumor segmentation. In this study, we propose a Deeply supervIsed knowledGE tranSfer neTwork (DIGEST), for brain tumor segmentation with incomplete multi-modal MRI data.
Our contributions are summarized as follows:

\begin{enumerate}
    \item To address the generalization limitations of the current deep learning methods in relying on the integrity of the multimodality information, this paper proposes DIGEST to learn distinguished features for accurate segmentation task empowered with teacher-student distillation framework. Specifically, training samples of the student network are generated by a Bernoulli sampling process, which helps elevate the student network‘s generalization ability for adapting the possible modality missing MRI data in real applications.
    \item To preserve the modality-shared information as far as possible and enhance the segmentation capability of the student network, a deeply supervised knowledge transfer loss is designed to constrain the consistency of the teacher-student structure of DIGEST at different decoding stages.
    \item Our method achieves promising results compared to two state-of-the-art methods on the widely utilized Brain Tumor Segmentation Challenge 2020 (BraTS 2020) dataset.
\end{enumerate}

\section{Method}
\subsection{Task formulation}
Given the MRI data $X$ containing the complete MRI modalities (T1, T1ce, T2, FLAIR), the ground truth segmentation map $Y$, and tumor region classes $C$, the multi-class segmentation task aims at finding a mapping $\mathcal{F}$ s.t. $\mathcal{F}(X)\rightarrow Y_{j}$, ${j}\in{C}$. For incomplete multi-modal brain tumor segmentation, the given MRI data become $\widetilde{X}_{k} \subseteq X,k \in K$ ($K$ refers to the number of possible combinations of incomplete MRI modalities), the straightforward solution is to find the corresponding $\widetilde{\mathcal{F}}_{k}$ s.t. $ \widetilde{\mathcal{F}}_{k}(\widetilde{X}_{k}) \rightarrow Y_{j}$ for all $k\in K$. However, finding $\mathcal{F}_{k}$ for all the different $k$ is difficult and inefficient.
Here, we hope to find a general mapping $\widetilde{\mathcal{F}}$ s.t. $\widetilde{\mathcal{F}}(\widetilde{X}_{k})\rightarrow Y_{j}$ for all ${j}\in{C}$ and $k\in K$. With such definition, a single $\widetilde{\mathcal{F}}$ can be utilized for different incomplete multi-modal MR image segmentation scenarios. Considering the given incomplete multi-modal data ${\widetilde{X}}_k\subseteq X$,the overall optimization problem can be defined as:

\begin{equation}
\mathop{arg\min}\limits_{{\widetilde{X}}_k\subseteq X,j\in C}{L({\widetilde{\mathcal{F}}}\ ({\widetilde{X}}_k),Y_j)+R({\widetilde{X}}_k)}
\end{equation}
where $L$ denotes the loss function, and $R$ is the regularization over the given data.

\subsection{Deeply Supervised Knowledge Transfer Network}
The overall framework of DIGEST is shown in Fig.1. The framework includes a teacher network $\mathcal{F}_t(\theta_t)$ and a student network $\mathcal{F}_s(\theta_s)$ and both of them adopt a modified 3d-Unet backbone proposed in\cite{henry2020brain}. Deep supervision\cite{wang2015training} is embed at each decoding stage of $\mathcal{F}_t(\theta_t)$ and $\mathcal{F}_s(\theta_s)$ after $1 \times1$ convolutions which help generate auxiliary segmentation maps at each stage. The knowledge transfer learning process consists of two steps: (1) Pretrain $\mathcal{F}_t$ with the complete multi-modal MRI data $X$. (2) Let the student model $\mathcal{F}_s$ inherit the information extracted by $\mathcal{F}_t$ and force $\mathcal{F}_s$ to extract more effective features when incomplete multi-modal MRI data are provided.

\subsection{Traning Data Settings}
The input training mini-batch ${\widetilde{x}}_i$ for $\mathcal{F}_s$ is derived from its complete multi-modal input counterpart $x_i$ at every training iteration $i$. Let $n$ denotes the number of available modalities, since $n$ is uncertain, we can simply assume that each modality follows a binomial distribution $B(1,0.5)$. Given the total modality numbers $n_t$, the modality mask $m_i$ for each iteration is decided by replacing $n_t-n$ modalities with zeros through Bernoulli sampling. The training data for $\mathcal{F}_s$ is then generated by multiplying the complete multi-modal input mini-batch $x_i \in X$ with the generated modality mask $m_i$. To further improve the generalization capability of  $\mathcal{F}_s$ , mixed attention CBAM\cite{woo2018cbam} blocks are introduced to the encoders of $\mathcal{F}_s$.

\subsection{Loss Function}
Base on the deep supervision structure, we design a deeply supervised knowledge transfer loss, which constrains the consistency of the modality-shared information flow between $\mathcal{F}_t$ and $\mathcal{F}_s$ across different decoding levels. We calculate the absolute difference between the auxiliary segmentation outputs of $\mathcal{F}_t$ and $\mathcal{F}_s$ at each decoding level. Such constraint can guide $\mathcal{F}_s$ to learn from $\mathcal{F}_t$ under different receptive fields and better preserve the common modality representations. Given a batch size of $N$, our deeply supervised transfer training loss function is defined as:

\begin{equation}
L_{ds}=\frac{1}{N}\sum_{z}\left|D_z\left(x_z\right)-{\widetilde{D}}_z\left({\widetilde{x}}_z\right)\right|
\end{equation}

where $z$ denotes the decoding stage index, $x_z$ and ${\widetilde{x}}_z$ are the input features,$D_z$ and ${\widetilde{D}}_z$ refer to the operations for generating the auxiliary segmentation map. We use Dice loss to constrain the final segmentation map of  $\mathcal{F}_s(\theta_s)$:

\begin{equation}
    L_{seg}=1-\frac{1}{N}\sum_{j}\frac{S_j\ast R_j+\varepsilon}{S_j^2+R_j^2+\varepsilon}
\end{equation}

where $S_j$ and $R_j$ refer to the output and ground truth label at $jth$ channel, $\varepsilon$ is a smoothing factor. The overall loss function for DIGEST is:

\begin{equation}
    {L=L}_{ds}+L_{seg}
\end{equation}

\section{Experiment}
\subsection{Dataset}
Extensive experiments are conducted over BraTS 2020\cite{menze2014multimodal}. In total, there are 369 cases, each containing data of four MRI modalities: T1, T1ce, T2, and FLAIR. The lesion area in each case is separated into the GD-enhancing tumor (ET), the peritumoral edema (ED), and the necrotic and non-enhancing tumor core (NCR/NET) with radiologists approved annotations. These areas can be further combined into three clustered sub-regions: the enhancing tumor (ET), the tumor core (TC=ET+NCR), and the whole tumor (WT=ET+NCR+ED). All the volumes have been co-registered to the same anatomical template, interpolated to $1\ {\rm mm}^3$ resolution, and skull-stripped.

\begin{table*}[t]
    \centering
    \scalebox{0.9}{
    \begin{tabular}{ccccccccccccc}
        \toprule [2pt]
            \multicolumn{4}{c}{\textbf{Modalities}}& \multicolumn{3}{c}{\textbf{Enhancing(ET)}}&  \multicolumn{3}{c}{\textbf{Core(TC)}}& \multicolumn{3}{c}{\textbf{Complete(WT)}} \\
        \midrule
             \makebox[0.025\textwidth][c]{\small{T1}}&\makebox[0.025\textwidth][c]{\small
             {T1ce}}&\makebox[0.025\textwidth][c]{\small{T2}}&\makebox[0.025\textwidth][c]{\small{FLAIR}}&U-HeMIS&U-HVED&DIGEST&U-HeMIS&U-HVED &DIGEST&U-HeMIS&U-HVED&DIGEST\\
        \midrule
            {\scriptsize{\CIRCLE}} & \scriptsize{\Circle} & \scriptsize{\Circle} & {\scriptsize{\Circle}} &13.1 & 22.5 & \textbf{35.1} & 38.9 & 53.8 &\textbf{54.5} & 58.8 & \textbf{83.5} & 68.6\\  
            \scriptsize{\Circle} & \scriptsize{\CIRCLE} & \scriptsize{\Circle} & \scriptsize{\Circle} & 66.2 & 66.3 & \textbf{78.3} & 68.3 & 70.7 &\textbf{83.3} & 65.2 & 65.4 & \textbf{74.4}\\ 
            \scriptsize{\Circle} & \scriptsize{\Circle} & \scriptsize{\CIRCLE} & \scriptsize{\Circle} &21.2 & 31.0 & \textbf{39.1} & 50.7 & 57.9 &\textbf{60.1} & 76.7 & \textbf{80.9} & 78.6\\ 
            \scriptsize{\Circle} & \scriptsize{\Circle} & \scriptsize{\Circle} & \scriptsize{\CIRCLE} &26.5 & 22.5 & \textbf{39.8} & 51.6 & 53.8 &\textbf{60.4} & 79.6 & 83.5 & \textbf{84.2}\\ 
            \scriptsize{\CIRCLE} & \scriptsize{\CIRCLE} & \scriptsize{\Circle} & \scriptsize{\Circle} &68.2 & 68.1 & \textbf{79.6} & 70.8 & 73.2 &\textbf{84.1} & 69.8 & 69.7 & \textbf{77.6}\\
            \scriptsize{\CIRCLE} & \scriptsize{\Circle} & \scriptsize{\CIRCLE} & \scriptsize{\Circle} &25.3 & 30.5 & \textbf{42.4} & 52.7 & 59.2 &\textbf{63.8} & 79.2 & 81.7 & \textbf{84.3}\\
            \scriptsize{\CIRCLE} & \scriptsize{\Circle} & \scriptsize{\Circle} & \scriptsize{\CIRCLE} &28.4 & 20.4 & \textbf{42.9} & 55.9 & 55.7 &\textbf{65.0} & 83.2 & 84.3 & \textbf{85.8}\\ 
            \scriptsize{\Circle} & \scriptsize{\CIRCLE} & \scriptsize{\CIRCLE} & \scriptsize{\Circle} &71.8 & 72.6 & \textbf{81.0} & 74.6 & 79.5 &\textbf{87.1} & 81.1 & 83.5 &\textbf{85.5} \\
            \scriptsize{\Circle} & \scriptsize{\CIRCLE} & \scriptsize{\CIRCLE} & \scriptsize{\Circle} &71.6 & 72.6 & \textbf{78.7} & 73.4 & 76.7 &\textbf{86.1} & 85.2 & 86.5 & \textbf{88.6}\\
            \scriptsize{\Circle} & \scriptsize{\Circle} & \scriptsize{\CIRCLE} & \scriptsize{\CIRCLE} &20.8 &34.6 & \textbf{44.6} &59.6  &62.7  &\textbf{67.8} &85.4  &87.6 & \textbf{87.8}\\
            \scriptsize{\CIRCLE} & \scriptsize{\CIRCLE} & \scriptsize{\CIRCLE} & \scriptsize{\Circle} &72.5 &72.5 & \textbf{82.3} &76.1  &79.7  &\textbf{86.1} &83.1  &84.0 & \textbf{86.0}\\
            \scriptsize{\CIRCLE} & \scriptsize{\CIRCLE} & \scriptsize{\Circle} & \scriptsize{\CIRCLE} &74.1 &72.7 & \textbf{77.7} &74.6  &77.0  &\textbf{85.6} &86.4  &86.7 & \textbf{88.5}\\
            \scriptsize{\Circle} & \scriptsize{\CIRCLE} & \scriptsize{\CIRCLE} & \scriptsize{\CIRCLE} &72.9 &73.7 & \textbf{80.3} &76.8  &80.1  &\textbf{87.2} &87.9  &89.2 & \textbf{89.7}\\
            \scriptsize{\CIRCLE} & \scriptsize{\Circle} & \scriptsize{\CIRCLE} & \scriptsize{\CIRCLE} &33.0	&32.7 &\textbf{45.5} &60.7 &63.4 &\textbf{67.5} &86.9 &88.2	&\textbf{88.8}\\
            \scriptsize{\CIRCLE}& \scriptsize{\CIRCLE} & \scriptsize{\CIRCLE} & \scriptsize{\CIRCLE} &73.4	&73.6 &\textbf{81.2} &77.1 &80.2 &\textbf{87.0} &88.5	&89.2 &\textbf{90.2}\\
        \midrule
            \multicolumn{4}{c}{mean} &
            49.3	&51.1	&\textbf{61.9}	&64.1	&68.2	&\textbf{75.0}	&79.8	&82.9	&\textbf{83.9}\\
        \bottomrule[2pt]
        \end{tabular}}
    \caption{Quantitative results(Dice $\%$) of different methods under possible modality missing conditions. {\scriptsize{$\CIRCLE$}} denotes the modality is available and {\scriptsize{$\Circle$}} denotes the modality is missing.}
    \label{tab:my_label1}
\end{table*}

\subsection{Implementation Details}
The experiments for our DIGEST are performed in Pytorch with a single 32GB NVIDIA Tesla V100. We randomly split the 369 cases into a training set (220 cases), a validation set (74 cases), and a test set (75 cases). For each case, Images are cropped by a smallest bounding box containing the whole brain area and then randomly cropped to 128x128x128 sub-volumes. Our model is trained for 200 epochs with a batch size of 1 using ranger optimizer\cite{yong2020gradient}. 
The initial learning rate is set to 1e-4 and cosine decay\cite{he2019bag} is used after 100 epochs.
The smoothing factor $\varepsilon$ in Dice loss is set to 1.

\subsection{Results}
We compare the results of the proposed DIGEST with two state-of-the-art methods, HeMIS[5] and U-HVED[6] by calculating the Dice scores of ET, TC, and WT. For HeMIS, we adopt the UNet-shaped model introduced in [6] named U-HeMIS. Table 1 shows that DIGEST generates better results than both methods under most of the experimental conditions for segmenting all the three tumor sub-regions. DIGEST achieves the best mean Dice score of 0.839 on segmenting WT averaged over all the 15 different multi-modal input combinations. Compared to the two state-of-the-art methods, DIGEST improves the mean Dice score by 0.068 on segmenting TC and by 0.108 on segmenting ET. Results of ablation studies of our methods are listed in Table 2.
\begin{table}[hbt]
    \scalebox{0.9}{
    \begin{tabular}{ccccc}
        \toprule[2pt]
            \textbf{$K_{p}$} & \textbf{$L_{ds}$} & \textbf{Enhancing (ET)} & \textbf{Core (TC)} &\textbf{Complete (WT)}\\
        \midrule
            \ding{56} & \ding{56} & 53.0 & 63.2 & 66.8 \\
            \ding{52} & \ding{56} & 60.3 & 72.1 & 82.6 \\
            \ding{52} & \ding{52} & 61.9 & 75.0 & 83.9 \\
        \bottomrule[2pt]
        \end{tabular}}
    \caption{Ablation results (Dice $\%$) of our method. $K_{p}$: knowledge transfer learning process. $L_{ds}$: deeply supervised loss. Scores are averaged over all the 15 experimental conditions.}
\end{table}

Example qualitative results are plotted in Fig. 2. It can be observed that the segmentation performance of the tumor core of the two comparison methods is largely affected when the data of the important modality T1ce are missing. On the other hand, the proposed DIGEST is able to produce a very intact outline of these heterogeneous tumor areas (ED,ET and NCR) even when T1ce data are not provided.

\begin{figure}[!htb]
    \centering
    \includegraphics[width=0.40\textwidth]{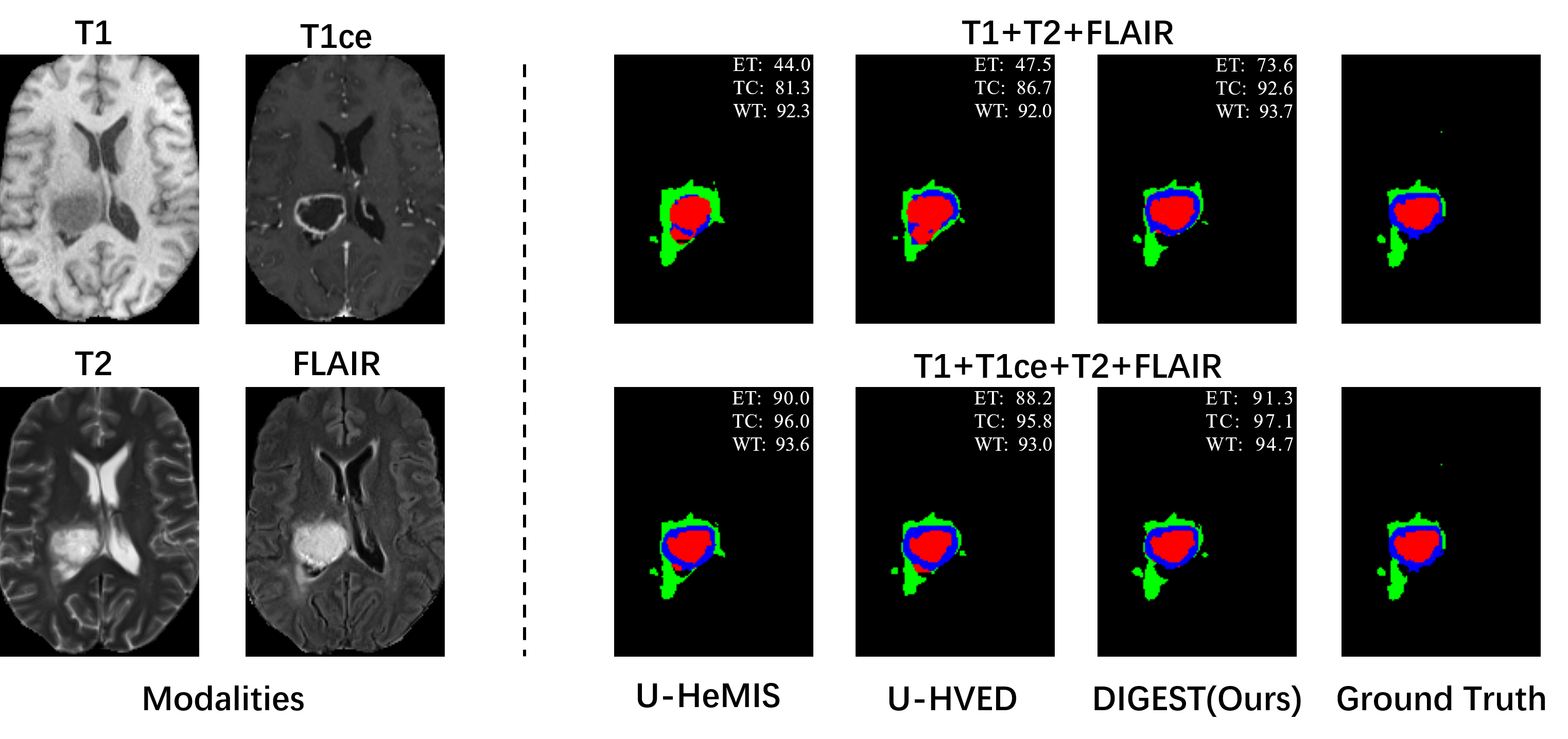}
    \caption{Example segmentation maps produced by the three methods. Green: ED. Blue: ET. Red: NCR. DIGEST provides a clear outline for TC (NCR and ET) even when the T1ce modality is missing.}
\end{figure}

\section{Conclusion}
In this study, we propose a deeply supervised knowledge transfer network learning model, DIGEST, for brain tumor segmentation when incomplete multi-modal MRI data are provided. DIGEST contains two important elements, a knowledge transfer learning network and a deeply supervised loss. Extensive experiments on the public dataset, BraTS 2020, have been conducted, and DIGEST generates promising brain tumor segmentation results when compared with two state-of-the-art methods. Owing to the flexibility of the input conditions, DIGEST has a high potential to be deployed in real-world clinical applications.

\section{Acknowledgement}
This research was partly supported by Scientific and Technical Innovation 2030-"New Generation Artificial Intelligence" Project (2020AAA0104100, 2020AAA0104105), the National Natural Science Foundation of China (61871371, 62222118, U22A2040), Guangdong Provincial Key Laboratory of Artificial Intelligence in Medical Image Analysis and Application (No. 2022B1212010011), the Basic Research Program of Shenzhen (JCYJ20180507182400762), Shenzhen Science and Technology Program (Grant No. RCYX20210706092104034), AND Youth Innovation Promotion Association Program of Chinese Academy of Sciences (2019351).

\bibliographystyle{IEEEbib}
\bibliography{strings,refs}

\end{document}